\documentclass{emulateapj}
\usepackage{graphicx}
\shortauthors{Jones, West, \& Foster}
\shorttitle{Mapping Extinction with M Dwarf Spectra}
\submitted{Accepted for Publication in AJ}
\begin{document}

\title{Using M Dwarf Spectra to Map Extinction in the Local Galaxy}
\author{David O. Jones\altaffilmark{1,2}, Andrew A. West \altaffilmark{2,3}, Jonathan B. Foster \altaffilmark{2}}

\altaffiltext{1}{Corresponding author: jonesd@bu.edu}
\altaffiltext{2}{Astronomy Department of Boston University, 725 Commonwealth Avenue, Boston, MA 02215, USA}
\altaffiltext{3}{Visiting Investigator, Department of Terrestrial Magnetism, Carnegie Institute of Washington, 5241 Broad Branch Road, NW, Washington, DC 20015, USA}

\begin{abstract} 

We use spectra of more than 56,000 M dwarfs from the Sloan Digital Sky Survey (SDSS) to create a high-latitude extinction map of the local Galaxy.  Our technique compares spectra from the stars in the SDSS Data Release 7 M dwarf sample in low-extinction lines of sight, as determined by Schlegel, Finkbeiner, \& Davis, to other SDSS M dwarf spectra in order to derive improved distance estimates and accurate line-of-sight extinctions.  Unlike most previous studies, which have used a two-color method to determine extinction, we fit extinction curves to fluxes across the spectral range from 5700 to 9200 \AA \ for every star in our sample.  Our result is an $A_V$ map that extends from a few tens of pc to approximately 2 kpc away from the sun.  We also use a similar technique to create a map of $R_V$ values within approximately 1 kpc of the sun, and find they are consistent with the widely accepted diffuse interstellar medium value of 3.1.  Using our extinction data, we derive a dust scale height for the local Galaxy of 119 $\pm$ 15 parsecs and find evidence for a local dust cavity.
\end{abstract}


\section{Introduction}

The distribution of Galactic dust is highly variable throughout the Galaxy, and despite many recent advances, its precise three-dimensional distribution is largely unknown.  Because dust both attenuates and reddens sources, the inability to properly correct for it places serious restrictions on the precision of many observational data products.  Dust is also important for star formation, since it is thought to be a catalyst for molecular gas formation and is often found in the cold, dense regions where new stars are born.

Currently the most widely used large-scale dust study is that of \citet*[hereafter SFD]{SFD98}, which created a two-dimensional map with total line-of-sight Milky Way dust column densities calculated from 100 and 240 $\mu$m emission.  \citet{AG99} determined that the SFD dust maps overestimate reddening by a factor of 1.3-1.5 in regions of smooth extinction with $A_V$ $>$ 0.5 and may underestimate reddening in regions of steep extinction gradients, and \citet{PG10} used the reddening of standard galaxies to find occasional deviations from SFD of up to 50\%.  In spite of these occasional inaccuracies, the SFD dust map generally allows observations outside the Galaxy to be corrected for the appropriate amount of dust.  However, sources inside the Galaxy lie in front of only a portion of the Milky Way's total dust and thus are likely attenuated and reddened by only a fraction of the total dust column.  This fraction varies based upon the 3D distribution of stars and dust in the Galaxy.

	Dust column density can be independently inferred from the total line of sight extinction to a given star.  The extinction affecting stellar spectra can be measured empirically by comparing intrinsic stellar colors to observed colors, using the formula:
\begin{equation}
A_V=R_V[E(B-V)], 
\end{equation}
where $R_V$ is the ratio of total extinction, $A_V$, to reddening, $E(B-V)$, and is an indicator of the dust grain size distribution and composition. $R_V$ has an average value of 3.1 for the diffuse interstellar medium, and ranges between about 2.1 and 5.5 for most sight lines in the Galaxy \citep*[hereafter CCM89]{CCM89}.  Higher values of $R_V$ are typical of dense regions (e.g. \citealp{R07}).

The relationship between total extinction and reddening can be described by an extinction law, usually a polynomial fit between the measured color excesses (e.g. $E(B-V)$) of different bands normalized to $A_V$.  The most commonly used extinction law is that of CCM89, which fits a law of the form $A(\lambda)/A_V=a(x)+b(x)/R_V$ to extinction measurements in Johnson photometric bands, where $a(x)$ and $b(x)$ are polynomial functions of x, which is in inverse wavelength units ($x=1/\lambda \ \mu m^{-1}$).  In CCM89 and other similar extinction laws (e.g. Rieke \& Lebofsky 1985; O'Donnell 1994; Fitzpatrick 1999), the shapes of extinction curves for optical and longer wavelengths depend only on the amount of dust and $R_V$, but their $R_V$-dependence varies considerably with wavelength.  Infrared curves depend only marginally on $R_V$ while shorter-wavelength optical curves have a strong $R_V$-dependence.
	
	Most of the extinction laws discussed above determine extinction by assuming that the reddening over each Johnson photometric band corresponds to reddening at the effective wavelength of the filter.  However, because reddening increases toward shorter wavelengths over a filter's width, the effective reddening wavelength is slightly shorter than the filter's effective wavelength.  By using modified effective reddening wavelengths as the basis for a spline fit to extinction values, \citet{F99} presents a reddening law that is more accurate when used with narrow-band photometry, other photometric systems, and spectroscopic data.
	
	In order to overcome the shortcomings of a two-dimensional extinction map (e.g. Burstein \& Heiles 1978, SFD), several attempts have been made to use color excesses and extinction laws to model Galactic extinction and dust in three dimensions.  \citet{H97} completed one of the most comprehensive all-sky empirical extinction models to date by combining the extinction surveys of \citet{F68}, \citet{NK80}, \citet{BP91}, and \citet{A92}.  Using Equation 1 (choosing an average $R_V$ of $\sim$3.1) in conjunction with intrinsic and observed absolute B and V magnitudes, \citet{H97} computed the line of sight extinctions for around 78,000 stars.

	These early models were important for providing an initial estimate of 3D extinction, but suffered from limited resolution and large measurement error due to unknown $R_V$ values.  The combined model of \citet{H97} was effective at identifying absorbing clouds within 5 kpc, but was only intended to identify large-scale extinction near the Galactic plane.  The studies that comprise the combined model were fairly limited, and even though the \citet{A92} study had the greatest number of stars, its accuracy was limited to about 1 kpc with a resolution of several degrees.  \citet{F68}, \citet{NK80}, and \citet{BP91} also had limited resolution due to the rare hot (mostly B-type) stars that comprised most of their samples.  There were also significant differences in data quality between the studies in the combined model, leading to poorly defined errors in many areas of the map.
	
	The assumption that $R_V$ equals 3.1 introduced further error in Hakkila et al. (1997) due to the short optical wavelengths used to compute extinction.  According to SFD, extreme $R_V$ values can cause deviations from the $R_V=3.1$ extinction law of up to 30\% in the $V$ band and 55\% in the $B$ band, so $E(B-V)$ could be subject to considerable error.  Although $R_V$ appears to be approximately 3.1 for most sight lines in the Galaxy, \citet{S10} and \citet{S10b} found a broad distribution of Galactic $R_V$ values.

	In recent years there have been considerable improvements to the \citet{H97} map at low Galactic latitudes, especially looking towards the Galactic center.  \citet{M06} created an extinction map using data from the Two Micron All Sky Survey (2MASS) for \textit{b} $\leq$ 10$^{\circ}$ and \textit{l} $\leq$ 100$^{\circ}$, an area that is often ignored in extinction maps.  The \citet{M06} model of the inner Galaxy contains high-resolution extinction values for a portion of the sky close to the Galactic plane.  They  compared $J-K_S$ colors to predicted intrinsic colors from the Besan\c con Model of the Galaxy \citep{R03} with K and M giants to determine the extinction in the inner Galaxy.  Because infrared extinction curves are assumed to be independent of $R_V$, this approach is more robust to $R_V$ variation than those using $B-V$ color excesses.  Similarly, \citet{S09} created an algorithm capable of determining extinction in A to early K-type stars by comparing intrinsic to observed $rÕ-iÕ$ colors (wavelengths with low $R_V$-dependence) from the INT/WFC Photometric H$\alpha$ Survey \citep{D05}, which they used to map extinction in high resolution at low Galactic latitudes.  Using a similar method as \citet{M06}, also with 2MASS photometry, \citet{G09,G10} used O-F type stars to map extinction within 1600 pc of the sun at a range of Galactic latitudes with greater resolution than is afforded by the more rare K and M giant stars.  Although higher than the resolution of \citet{M06} in local regions, the resolution of this map is currently limited to Galactocentric cartesian boxes of 100 pc per side.

        Photometric and spectroscopic data from the Sloan Digital Sky Survey \citep[SDSS]{Y00,A09} has recently been used to model reddening at higher latitudes.  \citet{S10} and \citet{S10b}, with photometry and spectroscopy, respectively, used the colors of main-sequence turn off stars to test reddening from the dust maps of SFD and to provide the first large-scale, two-dimensional map of $R_V$ in the Galaxy.  Because this survey needed to use bright stars with relatively invariant photometric colors, the stars used to calculate extinction were behind the entire dust column except at latitudes near the Galactic plane.

	Missing from these studies is a comprehensive, local, three-dimensional extinction map of higher Galactic latitudes.  M dwarfs, as the most ubiquitous stellar types in the Galaxy, are the most suitable candidates to pursue such a study. However, because of their low luminosity, M dwarfs have been almost completely ignored by previous extinction studies.  In addition, fitting extinction curves to photometric data for these stellar types (and many others) is much less robust than a spectroscopic method due to the fewer data points over which fitting can occur.  However, with the advent of large spectroscopic surveys such as the SDSS, there are now spectroscopic samples that contain more than 70,000 M dwarfs \citep{W11}.  Using these data, it is now possible to fit extinction curves to a large spectral range.  We now have the ability to pursue a more precise three-dimensional extinction map for high Galactic latitudes within about 2 kpc of the sun using M dwarfs.

        Using SDSS data to more precisely model extinction also gives us the ability to derive the three-dimensional distribution of dust in the Galaxy.  The main structural features of dust in the local Galaxy are the local bubble and an exponential disk.  The local bubble has been modeled in detail by measuring the lack of neutral gas within $\sim$50 to 200 pc of the sun.  Most recently, \citet{L03} modelled the local bubble in three dimensions by using absorption maps of NaI.  However, although the neutral gas profile of the local bubble is well-constrained, the dust column densities are not.  A local underdensity of reddening has been established (e.g. \citealp{K99}) and some of its contours have been shown to correlate with the gas cavity using reddening from stars with Hipparchos parallaxes \citep{V10}, but the precise shape and size of the dust cavity is not well known.  
       
        The scale height of dust in the local Galaxy is also relatively unconstrained.  The theoretical model of \citet{D01} finds a value $\sim$185 pc near the sun and the HI model (long thought to be a tracer of dust) of \citet{KK09} finds an approximate value of 150 pc at all Galactocentric radii, but these studies have a high uncertainty in the local environment.  Because M dwarfs are the most numerous population of stars and have been catalogued in great abundance in the nearby Galaxy \citep{B10}, they are ideal candidates to study the dust abundances in greater detail within several hundred pc of the sun.

	In this study, we used an SDSS spectroscopic sample of more than 56,000 M dwarfs to model Galactic dust and extinction at high latitudes in the local Galaxy.  In section 2 we discuss our spectroscopic sample, and in section 3 we discuss creating our extinction, $R_V$, and dust models.  Our results are presented in section 4 and discussed in section 5.

\section{Data}
        SDSS data have proven to be incredibly valuable for large-scale Galactic surveys due to their high quality photometry over large solid angles.  \citet{I08} used SDSS spectroscopy of 60,000 F and G-type main-sequence stars to model temperature and metallicity in the Galaxy.  \citet{J08} used photometric data from $\sim$48 million stars to create an unprecedented model of the stellar number density distribution.  SDSS has also been an ideal tool for studying the properties of low-mass stars. \citet{B10} used around 15 million low-mass dwarfs to measure the luminosity function and mass functions for these stars.  It also enabled large-scale studies of magnetic and chromospheric activity \citep{W04,W08}, and the use of low-mass stars as kinematic tracers of Galactic dynamics (\citealp{B07a,B11,F09}; Pineda et al. in preparation).

	Our spectroscopic sample was selected from the SDSS Data Release 7 (DR7; \citealp{A09}) M dwarf spectroscopic catalog \citep{W11}.  SDSS spectroscopy was carried out by twin fiber-fed spectrographs collecting 640 simultaneous observations.  Typical exposure times were $\sim$15-20 minutes, but exposures were subsequently co-added for exposure times of $\sim$45 minutes, producing medium resolution spectra with R $\sim$ 2000 \citep{Y00}.  SDSS spectroscopic plates each contained 16 spectrophotometric standard stars, which were selected by color to be F subdwarf stars.  The SDSS spectroscopic fluxes were calibrated by comparing the standard stars to a grid of theoretical spectra from Kurucz model atmospheres \citep{K92} and solving for a spectrophotometric solution on each plate.  The SDSS spectrophotometry matches the PSF photometry to 4\% rms \citep{A08}.  For more information on SDSS spectrophotometric calibration, see the discussions in the SDSS second and sixth data release papers \citep{A04,A08}.

        Our initial sample consisted of 70,823 M Dwarfs (M0 to M9), which were chosen based on typical M dwarf colors ($\textit{r}-\textit{i}>0.53$ and $\textit{i}-\textit{z}>0.3$; \citealp{W08}) and had spectral types verified by eye. To ensure the highest quality spectroscopic data, stars with signal-to-noise ratios near H$\alpha$ of less than 4 were removed.  Stars with colors indicating a possible M dwarf-White dwarf binary were also removed \citep{S04}, as were stars that didn't pass the SDSS photometric quality flags described in \citet{B10}.  The resulting sample contained 55,719 stars.
	
	The radial velocities were measured for all stars in the SDSS by cross-correlating high signal-to-noise ratio templates with each M dwarf spectrum \citep{B07b}.  Distances were measured for all stars using the photometric parallax technique described in \citet{B10}.  See \citet{W11} for more details on the sample selection and the value added quantities measured for the SDSS DR7 M dwarf sample.

\section{Analysis}

	In order to measure the alteration of stellar spectra by dust, we selected a set of template spectra from regions of the sky where the total integrated extinction (as determined by SFD) was less than 0.03 magnitudes in the \textit{r} band.  We then separated the total set of template spectra by spectral type.  For spectral types M7 and earlier, we also put spectra into 5 bins based on the strength of the TiO5 molecular band head.  The TiO bins were created to minimize the effects of metallicity on extinction curve fits (changes in metallicity can reduce the depth of the TiO molecular features; \citealp{G97,L07}).  The bins were computed so that each one had roughly the same number of template spectra, allowing every star to be compared against several templates.  Due to the weaker TiO5 features of M8 and M9 stars (caused by the condensation of TiO in their cooler atmospheres; \citealp{A95}) and the scarcity of low-extinction template stars for these types, we did not separate M8 and M9 templates into TiO5 bins.

	We visually inspected all of the template spectra for missing data or low signal-to-noise ratios over parts of their spectrum, finding that many of the stars in the original template sample were not suitable for use as ``ideal'' M dwarf spectra.  However, we managed to identify $\sim$5-10 extremely high quality spectra (out of $\sim$60 originally) for each TiO5 bin.   For the templates that we selected by eye, the median extinction given by SFD was 0.028 magnitudes in the \textit{r} band with a median distance of 141 pc.  Each of the template spectra were corrected for their radial velocity and spline-fitted to a wavelength array that was spaced 69.1 km s$^{-1}$ apart in velocity (SDSS wavelength spacing).

	Due to the low signal-to-noise ratios in the blue portion of many M dwarf spectra (caused by their red colors and low luminosities) and the high $R_V$-dependence of shorter-wavelength extinction laws, we limited our extinction fits to the region spanning from 5700 to 9200 \AA.  For our initial model we assumed $R_V$ was equal to 3.1 everywhere, which is a reasonable approximation for most lines of sight (CCM89).  Our assumed $R_V$ does not greatly increase our uncertainty because extinction at longer wavelengths, especially those longer than the $V$ band ($\sim$5500 \AA), has a relatively low $R_V$-dependence.  We derived a relationship between flux, distance, and extinction at each wavelength by using the fact that the ratio of fluxes between the program star and template star is a product of both the distance modulus and the extinction.  Using the $R_V$=3.1 extinction curve of \citet{F99}, we derived the following equation, which relates the ratio of fluxes between each star and low-extinction template star to the distance and total $V$ band extinction:
\begin{equation}
-2.5log\Big{[}{f(\lambda) \over{f_{t}(\lambda)}}\Big{]}=5log\Big{[}{d \over d_{t}}\Big{]}+A(\lambda),
\end{equation}
where $f$ is the flux of the program spectrum,  $f_{t}$ is the flux of the template spectrum, $d$ is the distance to the program star, $d_{t}$ is the distance to the template star, and $A(\lambda)$ is the extinction at wavelength $\lambda$ given by \citet{F99}.

	For each program M dwarf in the \citet{W11} sample, we corrected for radial velocity and interpolated the flux to the wavelength array of our template spectra.  Each program star flux array was divided by the flux array of every template star with the same spectral type and TiO5 strength.  This gave us values for the flux ratio at every wavelength.  The \citet{F99} extinction curves were parameterized by the IDL program {\ttfamily fm\_unred}\footnote[1]{Part of the IDL Astronomy User's Library at: http://idlastro.gsfc.nasa.gov/}, which we modified slightly to return the $A(\lambda)/E(B-V)$ extinction curve.  Dividing this curve by $R_V=A_V/E(B-V)$ gave us the relation normalized to $A(\lambda)/A_V$.  Multiplying the extinction curve by $A_V$ returned $A(\lambda)$, which can be described by the equation:

\begin{equation}
A(\lambda)=(A_V/R_V)[A(\lambda)/E(B-V)]_{Fitzpatrick},
\end{equation}

\noindent where $A_V$ was the only unknown.  

\begin{figure}
\plotone{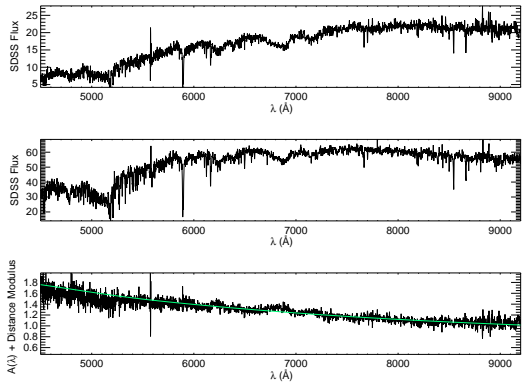}
\caption{Top: An M0 spectrum with some reddening.  Middle: An M0 template spectrum with less than 0.03 magnitudes of \textit{r} band reddening.  Bottom: 2.5 times the logarithm of the flux ratios of these two spectra, which is equal to the extinction added to the distance modulus.  This ratio has been fit with the \citet{F99} $R_V=3.1$ extinction law (green line).  The reddening is apparent in the slope of the extinction law fit, which had an $A_V$ of 0.88 magnitudes.  The SDSS flux is in units of $1\times10^{-17} erg \ s^{-1} \ cm^{-2} \ $\AA$^{-1}$.}
\label{figure:Ext Curve 2}
\end{figure}

Assuming that the derived distances \citep{B10,W11} were correct for the case of low-extinction templates, we were left with 2 free parameters, the program star distance, $d$, and the V band extinction, $A_V$.  We then performed a two-parameter Levenberg-Marquart least-squares fit using {\ttfamily mpfitfun} in IDL \citep{M09}, which returned the best fit $A_V$ and distance based on the fluxes at every wavelength (Figure \ref{figure:Ext Curve 2}).  Any residuals in the fits are likely due to noisy spectra or an imperfect match of program star to template star.  Taking the median of the fits for each template star gave the stellar distance and absolute V band extinction for each program star.  The dispersion in the values measured for each template star was used for the uncertainty.  The fits were constrained to have both distance and $A_V \ge 0.0$ and were weighted by the noise at each wavelength.  We were able to fit $A_V$ to a median uncertainty of 0.043 magnitudes and distance to a median uncertainty of less than 5\%.  This adds to the uncertainty in the original distances given by \citet{W11}, which is typically $\sim$20\%.  With the new distances, we derived a set of cylindrical Galactocentric coordinates for each star in the sample and have updated the sample distances in the \citet{W11} catalog.  To compute the error in $A_V$, we took the dispersion in $A_V$ for each program star and divided by the square root of the number of template stars against which each program star was fit.

\subsection{Fitting $R_V$}

\begin{figure}
\plotone{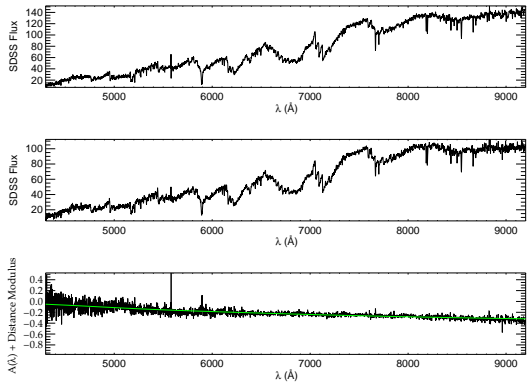}
\caption{Top: a M3 spectrum with some reddening.  Middle: a M3 template spectrum with less than 0.03 magnitudes of \textit{r} band reddening.  Bottom: 2.5 times the logarithm of the flux ratios of these two spectra, which is equal to the extinction added to the distance modulus.  This ratio has been fit with the \citet{F99} extinction law (green line; $A_V=0.81$; $R_V=3.31$).  The SDSS flux is in units of $1\times10^{-17} erg \ s^{-1} \ cm^{-2} \ $\AA$^{-1}$.}
\label{figure:Ext Curve 3}
\end{figure}

\begin{figure}
\plotone{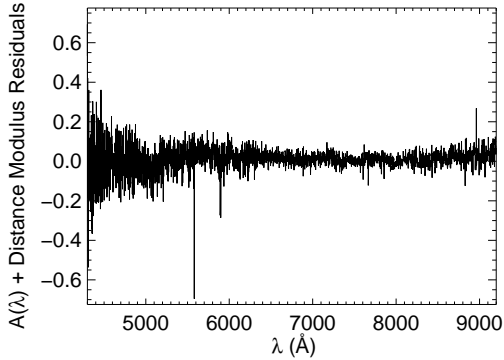}
\caption{A plot of the residuals for the extinction curve fit in Figure \ref{figure:Ext Curve 3}.  The residuals are generally small, with a median value 0.037, but tend to increase towards bluer wavelengths.}
\label{figure:Ext Residuals}
\end{figure}

	A subset of our sample was used to examine the distribution of $R_V$.  For the program and template stars with signal-to-noise ratios near H$\delta$ of greater than 4, we used the same procedure described above to perform a 3-parameter fit, this time including $R_V$ in Equation 3 as a free parameter.  Due to the high signal-to-noise ratio at the blue end of the spectrum for these stars, we were able to fit over the larger wavelength range of 4300 to 9200 \AA \ for 9102 M dwarfs (Figure \ref{figure:Ext Curve 3}).  As with our two-parameter fit, the residuals are likely due to an imperfect match of program spectrum to template spectrum.  A plot of the Figure \ref{figure:Ext Curve 3} extinction curve residuals is shown in Figure \ref{figure:Ext Residuals}.  The residuals shown are generally small, having a median value of 0.037, but tend to increase toward bluer wavelengths.  Unlike with the 2-parameter fits described above, where $A_V$ was restricted to have a minimum value of 0, we found that {\ttfamily mpfitfun} was better able to constrain $R_V$ when we allowed $A_V$ to be slightly negative.  We therefore allowed $A_V$ to be as low as -0.036 magnitudes, the systematic error from using a template star with 0.03 magnitudes of \textit{r} band extinction (assuming $R_V$ of 3.1), which allowed robust fits of $R_V$ that would be less affected by the systematic error in $A_V$.

        We found that $R_V$ is a sensitive parameter to the uncertainty in the spectrum, due to the fact that it describes the subtle changes in the curvature of extinction laws.  Many of the noisier fits returned values that were too high or too low to be plausible for Galactic sightlines.  We expect that these values, rather than being indicative of unlikely Galactic dust properties, could be due to the effect of random noise or possibly molecular features which could be slightly different in the program star than in the template star.  Our $R_V$ map also demonstrates that stars at greater distances, which are more likely to have lower-quality spectra, show the greatest dispersions in $R_V$, indicating that these values might not be physically meaningful.  In addition, the full width at half maximum of our distribution is $\sim$2, so these values appear to be well outside the normal range.  For these reasons, we constrained our analysis to $R_V$ values between 2.1 and 5.5 (CCM89).  Values in this range were returned for 6338 of the M dwarfs ($\sim$70\%) when compared with at least one template star.  The remaining 30\% of stars reported unlikely $R_V$ values when compared with all template stars.  The values between 2.1 and 5.5 were kept and used for a Galactic model of $R_V$ and to glean a rough sense of the overall distribution of $R_V$ values.

As with $A_V$, we computed the error in $R_V$ values by first taking the dispersion of the median for the $R_V$ values reported for each program star.  We then divided by the square root of the number of template stars against which each program star was fit.  However, we often found that only one template star for a given program star returned a realistic value of $R_V$.  For these stars, we used the formal 1-$\sigma$ errors reported by {\ttfamily mpfitfun}.  We were able to obtain a median uncertainty in $R_V$ of 0.42.

\subsection{Fitting to an Exponential Model}

In an effort to discover the large-scale distribution of local dust, we subsequently fit our extinction values to an integrated exponential dust model scaling in Galactocentric radius ($R$) and height ($Z$) and uniform with respect to the Galactocentric cylindrical angle $\theta$.  By doing this, we derived the approximate scale height and length of the dust.  Extinction is related to dust column density by the formula:

\begin{equation}
A_V({\bf{x}})=1.086\kappa_V \int_0^{\bf{x}} \! \rho_d (s) \,ds,
\end{equation}
\noindent
where $\kappa_V$ is the mean opacity of the dust \citep{Drimmel03}.  The mean opacity is estimated to be $8.55 \times 10^3$ $cm^2$$g^{-1}$ by the model of \citet{WD01} with slightly reduced abundances to match the canonical $A_V/N_H$ relation and optical constraints from \citet{Draine03,D03b}.  The scale height derived from this model, however, does not depend on which value of $\kappa_V$ we use.  We assumed an exponential dust model scaling in the Galactocentric $R$ and $Z$ coordinates of the form $\Sigma ({\bf{R,Z}})=\Sigma_0 \exp(-R/H_R)\exp(-Z/H_Z)$, which we integrated analytically to find the extinctions predicted by our model.  In order to derive the scale height of the dust, the integrated extinction was fit to the observed extinction by {\ttfamily mpfit2Dfun} \citep{M09} with $H_Z$, $H_R$, and $\Sigma_0$ as free parameters.  The corrected photometric parallax distances \citep{B10} used in this fit had a typical uncertainty of $\sim$25\%.

\subsection{Fitting to the Local Bubble}

Inspection of our $A_V$ model revealed a potential absence of dust near the Galactic plane that may be associated with the local bubble.  Although the local bubble has only been well-constrained by its absence of neutral gas, neutral gas tends to be a tracer for dust \citep{Draine03} and thus we expect the local bubble to contain a significant underdensity of dust.  We therefore decided to add the local bubble to our exponential dust model.  Although the shape of the local bubble is highly irregular, our limited resolution at small distances is only sufficient to roughly constrain the size of the bubble (due to the small number of nearby M dwarfs that don't saturate the SDSS detectors).  Because of this constraint and the approximate shape of the three-dimensional model of \citet{L03}, we decided to adopt a spherical model for the local bubble.  This model roughly matches the shape of the local bubble of ionized gas.  

Using the assumption of a spherical local bubble, we adopted an iterative approach wherein we used {\ttfamily mpfit2Dfun} \citep{M09} to fit to the scale height and scale length of an exponential model with a spherical cavity around the sun of varying radii.  Our free parameters were the scale height, the scale length, the central density ($\Sigma_0$), and the fraction of dust in the local bubble relative to the fraction predicted by our exponential disk model.  We first computed the reduced chi-square for bubble radii with the dust fraction parameter fixed at 0.0.  We adopted the bubble radius with the lowest chi-square value.  Next, we included the dust fraction as a free parameter in order to probe the density of dust within the local bubble.

\begin{figure}
\plotone{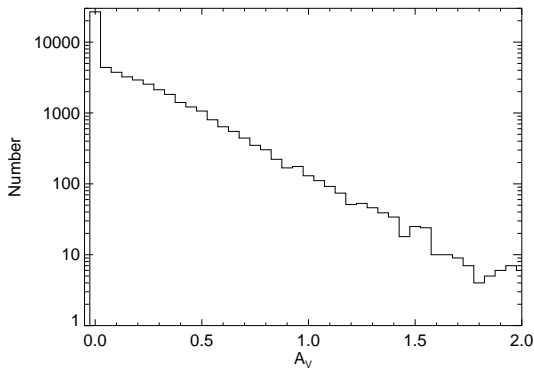}
\caption{A histogram of our values for $A_V$.  The values where $A_V=0.0$ comprise $\sim$40\% of our sample ($\sim$25,000 stars).  The distribution falls off exponentially toward higher $A_V$, which we would expect given that the histogram is dependent on the dust distribution, which we expect to scale exponentially with distance (from our analytical integration of Equation 4), and the sample density of M dwarfs as a function of distance.  The sampled M dwarf density begins to drop off exponentially after $\sim$200 pc, well before the median sampled M dwarf distance of $\sim$500 pc.}
\label{figure:AvHist}
\end{figure}

\begin{figure}
\plotone{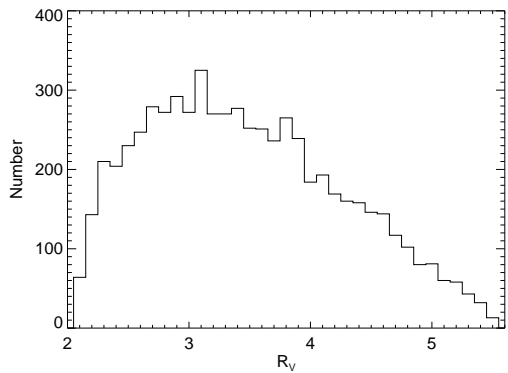}
\caption{A histogram of our $R_V$ values.  The distribution peaks close to the widely accepted diffuse interstellar medium value of 3.1, although our median  $R_V$ value of 3.38 is located slightly to the right of the peak.  Regions of higher dust column density are expected to account for the distribution's elongated tail.}
\label{figure:$R_V$ Histogram}
\end{figure}

\section{Results}

Using the methods described above, we measured $A_V$ and distance using 55,719 M dwarf spectra.  For more than 6,000 high signal-to-noise spectra, we also estimated $R_V$.  Histograms of our derived values for $A_V$ and $R_V$ are shown in Figures \ref{figure:AvHist} and \ref{figure:$R_V$ Histogram}, respectively.  Using these data, we created maps of $A_V$ and $R_V$, as well as a best fit model for the Galactic dust distribution.  We also compared our values to those of SFD for all lines of sight.  The details of our results are discussed below.

\subsection{Extinction Map}

\begin{figure*}
\plotone{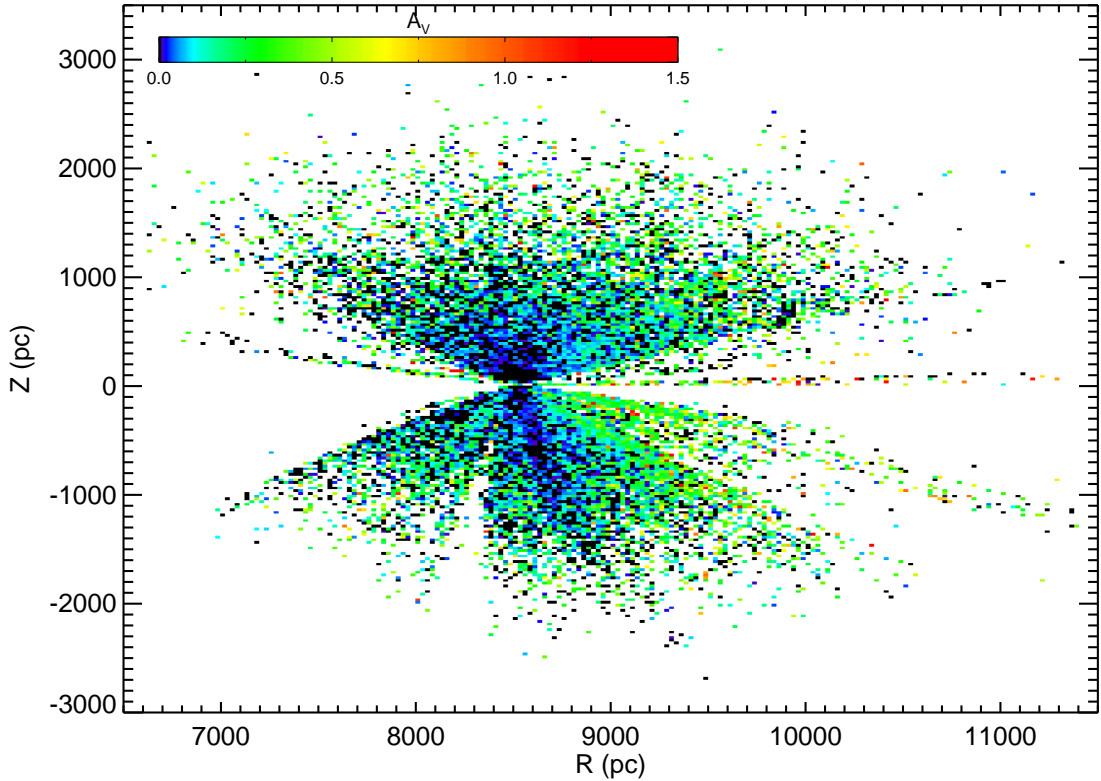}
\caption{Median $A_V$ (color coding) in Galactocentric cylindrical coordinate (R, Z) bins of 25 pc by 25 pc.  The map is flattened in the $\theta$-dimension.}
\label{figure:A(V)}
\end{figure*}

Figure \ref{figure:A(V)} shows the $A_V$ map from our analysis, created using the median $A_V$ values in 25 pc by 25 pc Galactocentric bins compressed in the $\theta$-dimension.  As these bins are in Galactocentric coordinates, their angular size varies with distance while their spatial size remains constant.  The extinction map extends to approximately 2 kpc above and below the Galactic plane.  In the northern hemisphere, the distribution of extinction across different lines of sight appears to be fairly regular, whereas in the southern hemisphere it appears patchier due to the relatively few lines of sight in the southern SDSS survey area (which likely does not represent the average line of sight extinctions in the southern Galactic hemisphere).  Although extinction generally increases at larger distances, there may be a selection effect at distances greater than around 1 kpc from the sun, where proportionally more of the stars observed are from low-extinction lines of sight.  This bias might arise because stars in high-extinction lines of sight are generally noisier and less likely to remain in our sample after our signal-to-noise cuts.

The median derived extinction for the entire sample is 0.037$\pm$0.029 magnitudes.  A histogram of our $A_V$ values (Figure \ref{figure:AvHist}) shows that the number of extinction values measured tends to decrease exponentially towards higher $A_V$.  This distribution is anticipated due to the fact that $A_V$ depends on the dust distribution, which scales exponentially along a line of sight, and the M dwarf sample density, which also scales exponentially in distance (outside of $\sim$200 pc).  This histogram also shows that 43\% of our sample reported an $A_V$ of 0.0.  Out of these values, 24\% reported no dispersion in their values due to the vast majority of the template star fits all reporting an $A_V$ of 0.0.  SFD report a median $A_V$ of 0.094 magnitudes for these stars, while they report a median $A_V$ of 0.097 for the entire M dwarf sample.  The stars to which we measure an $A_V$ of 0.0 have a median distance of 470 pc, well beyond the typical scale scale height of the dust disk, and so many of these values could partly be caused by random and systematic error or unusual lines of sight.  Out of the stars with a detecteable $A_V$, the median dispersion is 0.043 magnitudes.  We assumed that these dispersions represent the random errors in our extinction fitting; every program star was fit against multiple templates and the uncertainly can be gleaned from the deviation of all of the fits for a single program star from the median value.

In addition to the random error in our sample, there is also some systematic error that arises from the fact that our templates are not from perfectly extinction-free lines-of-sight.  Our template stars were chosen from regions of total column extinction of up to 0.03 magnitudes in the \textit{r} band.  Assuming an $R_V$ of 3.1, this translates to 0.036 magnitudes in $A_V$.  For the templates that we selected by eye however, the median total column extinction as given by SFD was 0.032 magnitudes in $A_V$, but with a median distance of only 141 pc.  Our results indicate that this median distance is only slightly above the average dust scale height of the Galaxy (see Section 4.3), and thus we expect that the actual $A_V$ to our templates is less than the maximum value.  The small median distances are due to the template spectra being selected to have high signal-to-noise ratios.  This process preferentially selects brighter stars, which are more likely to be at smaller distances and have $A_V$ values lower than the total line of sight values.  In addition, the difference in intrinsic template spectra extinction could contribute to the reported error in our fitting procedure.  Because template spectra have varying amounts of intrinsic $A_V$, the deviation in our template star fits will be in part due to the systematic error.  Because of this, our systematic error is likely less than the maximum value of 0.036 magnitudes.

\begin{figure}
\plotone{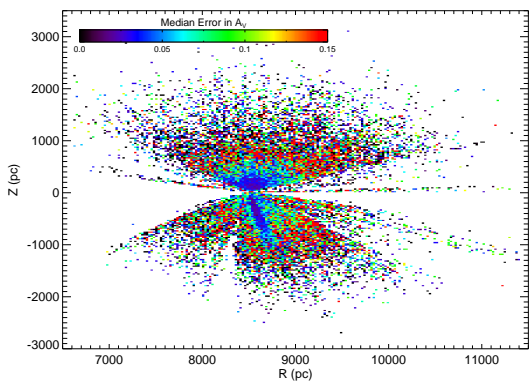}
\caption{Median error (color coding) of our $A_V$ map in Galactocentric cylindrical coordinate (R, Z) bins of 25 pc by 25 pc and flattened in the $\theta$-dimension.  Generally, the error in our values increases at greater distances.}
\label{figure:Error}
\end{figure}

Figure \ref{figure:Error} shows a map of the error in our $A_V$ map (Figure \ref{figure:A(V)}).  For the 25 pc by 25 pc Galactocentric bins in this map, we computed errors by taking the deviation from the median value of $A_V$ for all the stars in each bin and dividing by the square root of the number of stars.  For bins with only one star, we used the error measured by our fitting procedure.  As expected, the error in our values tends to increase at greater distances as spectra become noisier and less reliable.  For the 25 pc by 25 pc bins, the median error is 0.046.  The small increase in the median error when using these bins likely demonstrates the patchiness of the Galactic dust distribution (we have compressed all lines of sight in the $\theta$-dimension).

\begin{figure}
\plotone{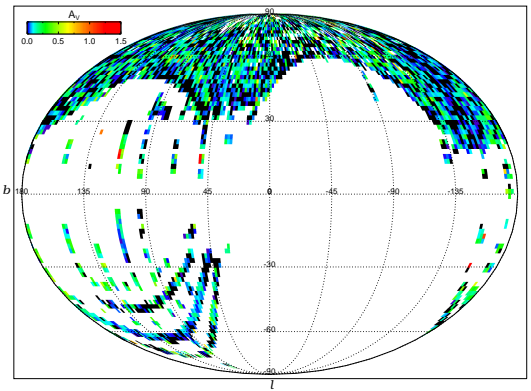}
\caption{A Galactic latitude and longitude map of our $A_V$ values (color coding) for all the stars in our sample averaged over bins of 4 square degrees.}
\label{figure:A(V) in Lat Lon}
\end{figure}

In addition to the map of $A_V$ in Galactocentric cylindrical coordinates, we also present a map of our sample in Galactic latitude and longitude.  Figure \ref{figure:A(V) in Lat Lon} shows a map of $A_V$ that includes all the stars in our sample averaged over bins of 4 square degrees.  This map shows that the SDSS data in the southern hemisphere come from only a few lines of sight, which explains why the extinction in the southern hemisphere of our Galactocentric map (Figure \ref{figure:A(V)}) is much more patchy than the extinction in the northern hemisphere.

\subsection{$R_V$ Map}

\begin{figure}
\plotone{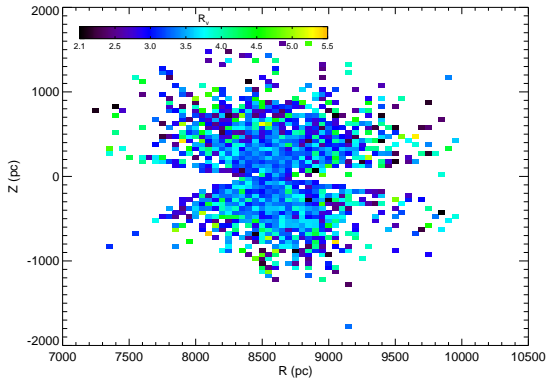}
\caption{Map of $R_V$ values (color coding) in Galactocentric cylindrical coordinate (R, Z) bins of 50 pc by 50 pc and flattened in the $\theta$-dimension.  The data give realistic values for $R_V$ to a distance of about 1 kpc.}
\label{figure:$R_V$}
\end{figure}

\begin{figure}
\plotone{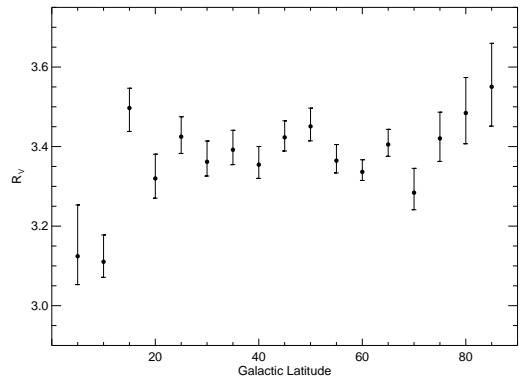}
\caption{Our derived $R_V$ values as a function of the absolute value of the Galactic latitude, $b$, in bins of 5$^{\circ}$.  It appears that there may be a correllation between $R_V$ and Galactic latitude. }
\label{figure:RvB}
\end{figure}

Using the 6,338 stars for which we measured realistic $R_V$ values, we created a map of our median $R_V$ for each 50 pc by 50 pc Galactocentric cylindrical coordinate bin (Figure \ref{figure:$R_V$}).  As with our $A_V$ map, the bins have an angular size which depends upon the distance at which they are located.  Accurate $R_V$ values proved challenging to determine at large distances.  Therefore our map of $R_V$ only extends to 1 kpc from the sun.  When possible, we measured the errors in $R_V$ by the same procedure described above for extinction values.  When only one template star returned a plausible value for $R_V$, we used the formal 1-$\sigma$ errors returned by {\ttfamily mpfitfun} \citep{M09}.  Figure \ref{figure:$R_V$ Histogram} shows the distribution of $R_V$, which peaks at a value of 3.1, in agreement with the generally accepted diffuse interstellar medium value (CCM89).  Our median $R_V$ is 3.38 with a median uncertainty of 0.42.  Figure \ref{figure:RvB} shows $R_V$ plotted as a function of Galactic latitude.  There appears to be an increase in $R_V$ values at higher Galactic latitudes, which is contrary to the expectation that $R_V$ is higher in the denser dust of the Galactic plane \citep{R07}.

We also examined the systematic error in our sample due to the fact that our template spectra have an unknown amount of reddening (up to 0.036 magnitudes in $A_V$).  Because of this, their intrinsic reddening contributes to the derived value of $R_V$.  We examined this error by using the IDL program {\ttfamily fm\_unred} to artificially redden our template spectra by an additional 0.036 magnitudes in $A_V$ using the \citet{F99} parameterization.  Our distribution, including the peak at an $R_V$ of 3.1 was mostly unchanged, but the median value shifted from 3.38 to 3.35 when the spectra were reddened.  Although we expect that our template spectra have less reddening than the maximum for the reasons discussed in section 4.1, there is still some systematic error.  Therefore, due to the unknown amount of intrinsic reddening of our template spectra, the true median $R_V$ could be slightly higher than the value we measured.

We next examined whether the observed distribution of $R_V$ was consistent with a Gaussian with the width equal to our uncertainty in $R_V$.  To do this, we created a random Gaussian distribution with 9,102 values of $R_V$ (our initial sample size) centered at 3.1 and with $\sigma=0.42$ (our median error).  We removed all values that were less than 2.1 or greater than 5.5.  We performed a Kolmogorov-Smirnoff test on the observed and simulated $R_V$ distributions, and found a 0\% probability that the two datasets were drawn from the same parent sample.  The reported distribution of $R_V$ is therefore broader than can be explained by a single Gaussian distribution with $\sigma=0.42$, and thus is indicative of a real spread in Galactic $R_V$.  However, some of the spread is likely due to our measurement uncertainties.

\subsection{Fitting to a Simple Dust Model}

To fit an exponential dust model with a local bubble to our extinction results, we first reduced the measured uncertainties by putting the stars into Galactocentric cylindrical (R, Z, R$\sin\theta$) three-dimensional boxes of 25 pc on a side at all distances.  We weighted each bin by the number of stars it contained and only used the stars within 1 kpc of the sun (where the dust dominates).

Using the iterative approach outlined in section 3, we looked for local minima in the reduced chi-square for a number of different local bubble sizes ranging from 0 pc to 350 pc in steps of 25 pc.  We found a best fit local bubble with a radius of 150 pc containing no dust.  We later included the dust fraction in the local bubble as a free parameter in our model and found the best fit dust fraction in a 150 pc radius was 0.40, i.e. the local bubble contains 40\% of the dust predicted by our exponential model.  Because the local bubble’s shape is highly irregular, this procedure is only meant to roughly constrain the local underdensity of dust.  Accordingly, the formal 1-$\sigma$ error in this fit is 26\%.  We performed an F test on our results to test whether the 40\% dust bubble is statistically more likely than the 0\% dust bubble, and found a significance of 0\%.  This result indicates that the addition of a parameter to constrain the dust fraction inside the bubble, although resulting in a lower chi-squared value, has no statistical significance.  Our model therefore finds that a 0\% dust bubble is as likely as a 40\% dust bubble.

The fitting procedure returned a dust scale height of 152$\pm$33 pc.  However, the small amount of data at low latitudes and our treatment of the dust structure inside our “spherical” local bubble limits the precision of our method.  As an alternative, we removed all stars with $A_V=0$ from the dataset before binning and fitting for the dust scale height, because we would not expect lines of sight without dust to scale exponentially.  This method effectively removed any effects of the zero extinction lines of sight from our derived scale height.  From this procedure, we found a dust scale height of 119$\pm$15 pc.  The actual scale height is likely bracketed by the results from our two methods.  Our derived scale length was nearly infinite for the first method, but after removing all stars with $A_V$=0, we found a value of 1.35$\pm$5.25 kpc.  The large uncertainty in this parameter is due to a lack of high-quality spectra at significantly different values of Galactocentric radii.

\begin{figure}
\plotone{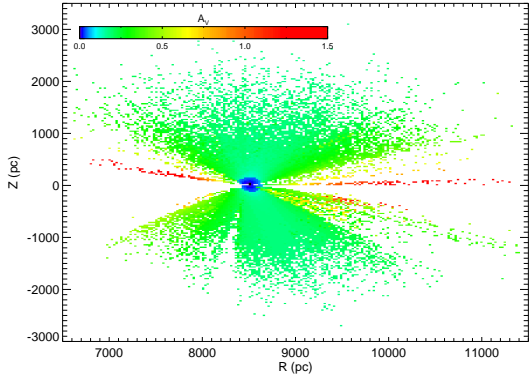}
\caption{The theoretical $A_V$ (color coding) assuming an exponential dust model that scales both in R and Z with a spherical local bubble that also scales exponentially but with 40\% of the dust density.  We can see from this model compared to our actual data (Figure \ref{figure:A(V)}) that the dust is considerably more patchy than a uniform exponential scaling.}
\label{figure:A(V) Exponential Model}
\end{figure}

\begin{figure}
\plotone{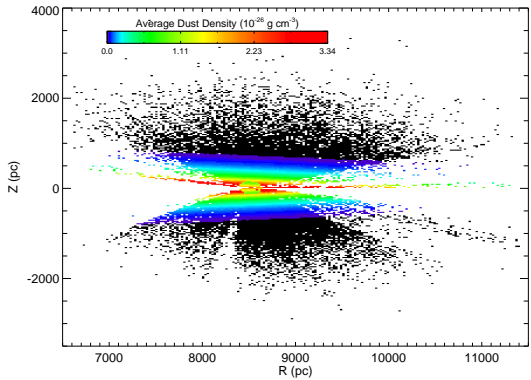}
\caption{The dust densities from our theoretical model (color coding), which finds a dust scale height of $119 \pm 15$ pc, a scale length of 1.35$\pm$5.25 kpc, and a spherical bubble with a radius of 150 pc containing 40\% of the dust predicted by a smooth exponential model.}
\label{figure:Dust Exponential Model}
\end{figure}

\begin{deluxetable}{lcc}
\tablewidth{0pt}
\tablewidth{0pt}
\tablecolumns{3} 
\tabletypesize{\scriptsize}
\tablecaption{Best-Fit Dust Parameters}
\renewcommand{\arraystretch}{.6}
\tablehead{
\colhead{Parameter}&
\colhead{Value}}
\startdata
Central Density&47 $\pm$ 117 $\times$ $10^{-26}$ g cm$^{-3}$\\
Scale Height&119 $\pm$ 15 pc\\
Scale Length&1.35 $\pm$ 5.25 kpc\\
Local Bubble Radius&150 pc\\
Local Bubble Dust Fraction&40 $\pm$ 26\%
\enddata
\label{table:dustparams}
\end{deluxetable}

\begin{figure}
\plotone{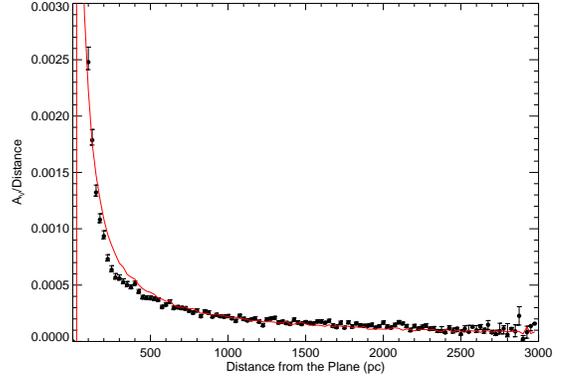}
\caption{Median $A_V$ divided by distance, an indicator of dust, as a function of distance from the plane in bins of 25 pc.  We have left out the values of $A_V=0$, because we don't expect lines of sight with no detectable dust to scale exponentially.  However, this plot shows that the lines of sight with significant dust tend to scale exponentially.  Our best fit extinction model is plotted in red and closely matches the data.  Note that the best fit extinction model is a fit to our extinction map and is not a fit to the data shown.}
\label{figure:ExtvDist}
\end{figure}

Figures \ref{figure:A(V) Exponential Model} and \ref{figure:Dust Exponential Model} show the theoretical extinction and dust distributions, respectively, using the scale length and height from the second method above and the local bubble size and dust fraction from the first method.  Table \ref{table:dustparams} contains the list of best-fit dust parameters used in these models.  The actual extinction does vary considerably compared to the uniformly scaling best-fit model, which indicates the patchiness of the dust distribution.  Figure \ref{figure:ExtvDist} shows $A_V$ divided by distance for the stars in our sample with $A_V$ greater than 0.  $A_V$ divided by distance is an indicator of dust because dust is proportional to the differential extinction.  This means that extinction divided by the distance is directly proportional to the average dust density along the entire line of sight to the star (see Equation 4).  Figure \ref{figure:ExtvDist} shows that the dust does appear to scale exponentially, on average, and fits our derived exponential extinction model very closely.

\subsection{Comparison with SFD}

\begin{figure*}
\plotone{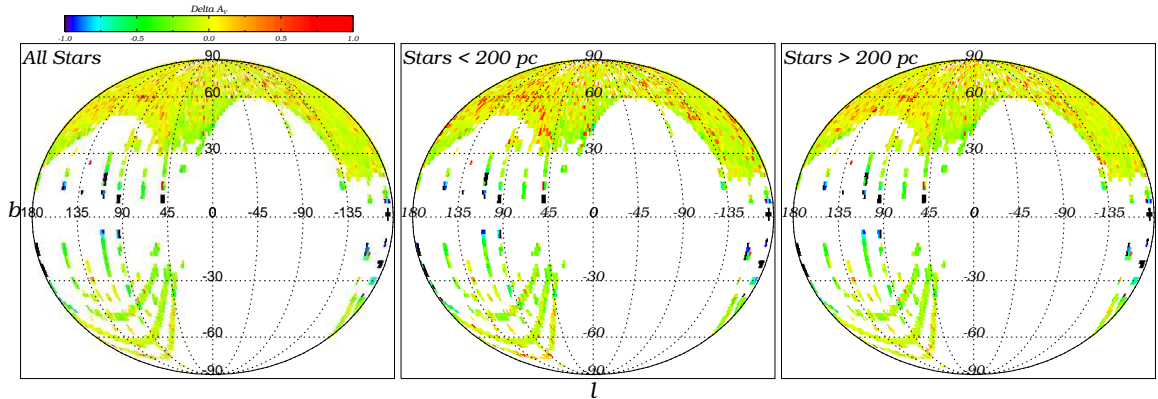}
\caption{A Galactic latitude and longitude map of the values of $A_V$ determined by SFD subtracted from our values of $A_V$ (color coding) for all the stars in our sample (left), stars with distance $<$ 200 pc (center), and stars with distance $>$ 200 pc (right).  In this plot we assume an $R_V$=3.1 extinction law.  Many lines of sight have significantly lower extinction than the SFD values, suggesting that these stars are in front of a significant portion of the dust column.  We expect that the places where our values are lower than those of SFD at low latitudes are due to the fact that we are not seeing through all of the dust near the plane.  This is consistent with the findings of \citet{S10} and \citet{S10b}.  In the center map (distance $<$ 200 pc) there are a few more lines of sight where our values are lower than those of SFD, suggesting that these stars are in front of a significant portion of the dust column.  There is an interesting feature in the upper left, where our values are higher than those of SFD at a variety of distances, which will be examined in more detail in future studies.  In the map on the right (distance $>$ 200 pc), there are few lines of sight where we find lower values than those of SFD, suggesting that many of these stars are beyond one dust scale height.}
\label{figure:SFD Comparison}
\end{figure*}

Figure \ref{figure:SFD Comparison} shows a map in Galactic latitude and longitude of SFD's total column extinction subtracted from our $R_V=3.1$ $A_V$ values.  Although our values were usually in close agreement to those of SFD, our values are often lower than the SFD values by a significant amount at lower Galactic latitudes.  This is expected due to the fact that we are not seeing through all of the dust near the Galactic plane, and our results for these regions are consistent with those of \citet{S10} and \citet{S10b}.  Areas where our results are lower than those of SFD at higher latitudes are also likely due to stars that are in front of a significant fraction of the dust.  Areas at higher latitudes that report higher $A_V$ values than those of SFD values could be due to error in our extinction values, our assumption of $R_V$=3.1, or possibly due to steep extinction gradients, where SFD may underestimate the extinction \citep{AG99}.  Representative of this, Figure \ref{figure:SFD Comparison} also shows comparisons with SFD for stars at distances less than and greater than 200 pc.  These maps show that our extinction values are more likely to be closer to those of SFD at greater distances.  In the upper left of the middle map (distance $<$ 200 pc) and in a few other lines of sight, however, we find values significantly higher than those of SFD.  We will examine these features in more detail in future studies.

\section{Conclusions}

By fitting over 56,000 SDSS M dwarf spectra to the \citet{F99} extinction curve, we have created extinction and $R_V$ maps for the local Galaxy.  Our results are summarized below.
\begin{enumerate}
\item We developed a technique for empirically measuring the extinction and distance to stars for which we have SDSS spectroscopic data.  Most previous techniques have used a two-color analysis to determine extinction; our method uses the flux values across the spectral range from 5700 to 9200 \AA \ to determine the best possible extinction and distance fits, and so has the potential to be much more accurate.
\item Using our fitting procedure, we derived new distances for most of the stars in the \citet{W11} SDSS DR7 M dwarf catalog, accounting for the proper amount of foreground dust extinction.
\item We created an $A_V$ map extending to approximately 2 kpc from the sun, generally at high Galactic latitudes.   In the northern hemisphere, where most of the stars in the SDSS sample lie, the $\theta$-compressed extinction is relatively uniform.  In the southern hemisphere, however, there are fewer SDSS lines of sight and so significant variations are apparent, showing that the three-dimensional dust distribution often does not scale uniformly.
\item Our extinction map compares favorably to the SFD map over the vast majority of the SDSS survey area.  Our extinction values are often significantly less than those of SFD, most likely because we are seeing through only part of the total dust column. The median error in our extinction values is 0.029 magnitudes.
\item Within about 1 kpc of the sun, we deduced the value of $R_V$ for many lines of sight, finding that the median $R_V$ is 3.38 with a median uncertainty of 0.42.  The peak of our distribution of $R_V$ (Figure \ref{figure:$R_V$ Histogram}) is at a value of 3.1, in agreement with the generally accepted value for the diffuse interstellar medium.  As we find no significant source of systematic error and a K-S test indicates that this distribution is inconsistent with a Gaussian of $\sigma=0.42$, we expect that the spread in our distribution indicates a real spread in Galactic $R_V$.
\item We estimated the scale height of dust in the local Galaxy to be 119 $\pm$ 15 parsecs.  This is less than the value found by the theoretical model of \citet{D01} which finds a value around 185 pc near the sun and roughly consistent with the HI model (long thought to be a tracer of dust) of \citet{KK09} which finds an approximate value at all galactocentric radii of 150 pc.  We note that there is considerable variation in the structure of dust throughout the local Galaxy.  In addition, we find some evidence for a spherical local bubble of radius 150 pc and with approximately 40\% of the normal dust density.  We plan to better constrain the shape, size and dust content of the local bubble through future work using nearby L dwarfs.  The empirical data varies considerably compared our model, but on average it is in close agreement (Figure \ref{figure:ExtvDist}). Similarly, the true dust distribution is not nearly as uniform as our model.
\end{enumerate}

Our 3D map of $A_V$ is most appropriate for estimating extinction to stars within around 500 parsecs of the Galactic plane.  Our map within this region, especially at small distances, has fairly high-resolution and low uncertainties, and using the SFD maps for these Galactocentric heights could significantly overcorrect for extinction.  Due to an approximate (but not uniform) dust scale height of 119 parsecs, we can safely assume that stars further than 500 parsecs from the plane are behind essentially all of the dust column and that two-dimensional maps like SFD's higher resolution map are the most appropriate for these distances.  

In using our map of $R_V$, it is important to remember that our standard error tends to be around 0.42; however, regions of especially high and low $R_V$ should be noted, especially for shorter wavelength observations where errors caused by assuming an average $R_V$ can be significant.  We expect that although for many lines of sight in the Galaxy $R_V=3.1$ is a very good approximation, there is a broad spread in Galactic $R_V$.  Because our $R_V$ histogram reports few $R_V$ values near 2.1 or 5.5, we do not expect that Galactic $R_V$ values are frequently outside of the range that we selected.  However, when our selection effect is taken into account, the spread in Galactic $R_V$ is even greater than that shown in our histogram.

 In the future, we plan to expand our extinction map by using the fitting procedure developed here for SDSS spectroscopic data from other stellar types.  Our complete $A_V$, $R_V$, and distance dataset is accessible online in FITS format at \textit{http://people.bu.edu/aawest/dust.html}.

\section{Acknowledgments}
We would like to thank the anonymous referee for many helpful suggestions.  We would also like to thank Eddie Schlafly, Douglas Finkbeiner, Fred Walter, Kevin Covey, John Bochanski, Alan Whiting, Tim Cook, Dan Clemens, and Josh Peek for useful discussions while conducting this study.

Funding for the Sloan Digital Sky Survey (SDSS) and SDSS-II has been
provided by the Alfred P. Sloan Foundation, the Participating
Institutions, the National Science Foundation, the U.S. Department of
Energy, the National Aeronautics and Space Administration, the
Japanese Monbukagakusho, and the Max Planck Society, and the Higher
Education Funding Council for England. The SDSS Web site is
http://www.sdss.org/.

The SDSS is managed by the Astrophysical Research Consortium (ARC) for
the Participating Institutions. The Participating Institutions are the
American Museum of Natural History, Astrophysical Institute Potsdam,
University of Basel, University of Cambridge, Case Western Reserve
University, The University of Chicago, Drexel University, Fermilab,
the Institute for Advanced Study, the Japan Participation Group, The
Johns Hopkins University, the Joint Institute for Nuclear
Astrophysics, the Kavli Institute for Particle Astrophysics and
Cosmology, the Korean Scientist Group, the Chinese Academy of Sciences
(LAMOST), Los Alamos National Laboratory, the Max-Planck-Institute for
Astronomy (MPIA), the Max-Planck-Institute for Astrophysics (MPA), New
Mexico State University, Ohio State University, University of
Pittsburgh, University of Portsmouth, Princeton University, the United
States Naval Observatory, and the University of Washington.

\bibliographystyle{aj}

\end{document}